\documentclass[aps,prl,twocolumn,showpacs,superscriptaddress,groupedaddress]{revtex4}
\usepackage{graphicx}
\usepackage{bm}        
\usepackage{amssymb}   
\begin{document}

\title{Free Energy Barrier for Electric Field Driven Polymer Entry into Nanoscale Channels }

\author{Narges Nikoofard}
\affiliation{Department of Physics, Institute for Advanced
Studies in Basic Sciences (IASBS), Zanjan 45137-66731, Iran}

\author{Hossein Fazli}
\email{fazli@iasbs.ac.ir} \affiliation{Department of Physics,
Institute for Advanced Studies in Basic Sciences (IASBS), Zanjan
45137-66731, Iran} \affiliation{Department of Biological Sciences,
Institute for Advanced Studies in Basic Sciences (IASBS), Zanjan
45137-66731, Iran}

\date{\today}

\begin{abstract}
Free energy barrier for entry of a charged polymer into a
nanoscale channel by a driving electric field is studied
theoretically and using molecular dynamics simulations.
Dependence of the barrier height on the polymer length, the
driving field strength, and the channel entrance geometry is
investigated. Squeezing effect of the electric field on the
polymer before its entry to the channel is taken into account. It
is shown that lateral confinement of the polymer prior to its
entry changes the polymer length dependence of the barrier height
noticeably. Our theory and simulation results are in good
agreement and reasonably describe related experimental data.

\end{abstract}

\maketitle

Interaction of polymer chains with nanoscale channels is a
phenomenon of rich basic science and numerous potential
applications
\cite{nanochannel,sequencing,confined,craighead1,separation,protein,2structure}.
Better understanding of related biological processes, analyzing
biopolymers like DNA \cite{sequencing}, investigating the
existing theories on static and dynamic properties of confined
polymers \cite{confined}, and separation of polymers of different
lengths \cite{craighead1,separation} are sample advantages of the
experiments in which polymers are forced to interact with
nanochannels. Also, this phenomenon provides possibility of
observation and direct study of protein-nucleic acid interactions
\cite{protein} and nucleic acids secondary structures
\cite{2structure}. A polymer entering into a narrow channel
experiences an entropic barrier due to the reduction of its
conformations and an external force should be applied
to overcome this barrier. A voltage difference in the case of
charged polymers like DNA \cite{voltage}, or flow injection may
provide the driving force \cite{pressure,pressure2}.

The entry process of a charged polymer into a nanochannel in a
wall under a voltage difference can be divided into two stages
\cite{capture}. First, the polymer comes to the channel vicinity
by a pure or electrically biased diffusion and one of its ends finds the channel
entrance. Then, if the driving field overcomes the entropic
barrier, the polymer enters the channel. Calculation of the
\begin{figure}
\includegraphics[scale=0.3]{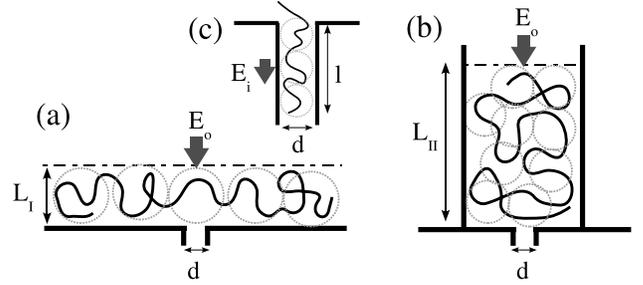}
\caption{\label{fig:epsart} Schematic of a polymer under the
electric field of strength $E_o$ behind a channel of diameter $d$
in a wall. (a) The polymer can be viewed as a 2D chain of blobs of
size $L_I$. (b) A polymer which is also confined laterally by a
cylinder of diameter $D$. The polymer can be considered as a
chain of blobs closely packed inside a cylinder of height
$L_{II}$. (c) A segment of the polymer entered to the channel
inside which the electric field is of strength $E_i$. }
\label{fig1}
\end{figure}
barrier height as a function of the system parameters such as the
driving field strength, the channel geometry, and properties of
the polymer is of great importance for numerous related
experiments. Polymer translocation through nanochannels and
polymer escape from entropic traps which recently has been used
as a polymer separation method \cite{craighead1,separation} are
examples of such experiments. Despite extensive experimental and theoretical
studies of driven polymer threading into nanoscale channels and
numerous suggestions for enhancing the polymer capture rate
\cite{capture,capture2,Chou-Grosberg,Muthukumar}, the entry barrier height and
its dependence on the system variables is less investigated.

In this paper, we calculate free energy barrier experienced by a
flexible charged polymer entering into a nanochannel under an
applied electric field theoretically and using coarse grained
molecular dynamics (MD) simulations. We obtain dependence of the
barrier height on the polymer length and the driving field
strength for two different geometries of the channel entrance.
The external driving field could have squeezing effect on the
polymer before its entry to the channel and affect its static and
dynamic behaviors. Our study shows that this squeezing effect and
lateral confinement of the polymer prior to the channel entry
noticeably affect the dependence of the barrier height on the
polymer length and the field strength. Our theoretical and MD
simulation results support each other and are in reasonable
agreement with some related experimental data.
\begin{figure*}
\includegraphics[scale=0.5]{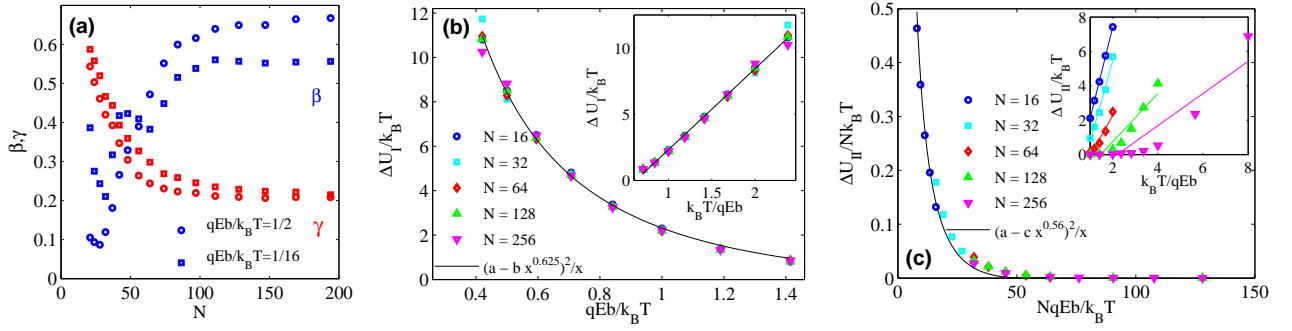}
\caption{\label{fig:epsart} (color online) (a) $\beta$ in
$L\propto N^{\beta}$ and $\gamma$ in
$R_{\parallel}L^{\frac{1}{4}}\propto N^{\gamma}$ versus the
polymer length. With increasing $N$ crossover from values
corresponding to case $I$ to those of case $II$ can be seen.
(b),(c) Simulation results for polymer entry barrier and fitted
functions of the theory for cases $I$ and $II$. Coefficients
a=3.02, b=1.50 and c=0.32 are fitting parameters. Considering
scaled axis of these figures, agreement between simulation and
theory can be seen. Insets: entry barriers versus $E^{-1}$ and
fitted lines.}\label{fig2}
\end{figure*}

To obtain free energy barrier for two channel entrance geometries
shown in Fig. \ref{fig1}, we calculate the polymer free energy
before its entry to the channel and after entering a fraction of
its monomers. We assume that electric fields inside and outside of
the channel are uniform and their strength may be different. As
case $I$, consider the polymer which contains $N$ charged
monomers of diameter $b$ and charge $q$, pushed to a wall by the
electric field of strength $E_o$ (see Fig. \ref{fig1} (a)). We
assume that there is no electrostatic interaction between monomers
(corresponding to high salt concentration) and that the monomers
only interact with the electric field. The polymer decreases its
electric energy by spreading on the wall in price of loosing a
part of its entropy. The polymer layer thickness, $L_I$, can be
determined by balancing entropic and electric energies. In this
layer, the polymer can be viewed as a 2D chain of blobs of size
$L_I$, where the number of monomers in each blob is $g_I \sim
\left(\frac{L_I}{b}\right)^{\frac{1}{\nu}}$ and $\nu$ is the Flory
exponent. Entropic energy of the polymer is $k_BT$ per blob,
$F_{ent} \sim k_BTN\left(\frac{b}{L_I}\right)^{\frac{1}{\nu}}$.
For case $II$, consider a similar polymer which is also confined
from sides by a cylinder of diameter $D$ (see Fig. \ref{fig1}
(b)). Because of simultaneous confining effects of the electric
field and the cylinder, this is similar to a polymer in a closed
space which its blobs are closely packed. The number of monomers
in each blob is given by
$g_{II}\sim\left(\frac{\Omega}{Nb^3}\right)^{\frac{1}{3\nu-1}}$,
where $\Omega \sim L_{II}D^2$ is the confinement volume. The
entropic energy of the polymer in this case is $F_{ent} \sim
k_BT\left(\frac{N^{3\nu}b^3}{\Omega}\right)^{\frac{1}{3\nu-1}} $
\cite{pressure}.

Electric energy of the polymer layer, $F_{elc}$, in both of these cases with
its zero taken on the wall is $\sim NqE_oL_m$ ($m=I,II$). The
equilibrium layer thickness can be obtained by
minimizing total free energy $F=F_{ent}+F_{elc}$ with respect to
$L_m$:
\begin{equation}  \label{L}
L_I \sim b\left(\frac{qE_ob}{k_BT}\right)^{\frac{-\nu}{1+\nu}},  \\
L_{II} \sim b\left(\frac{qE_ob}{k_BT}\right)^{\frac{1-3\nu}{3\nu}}
\left(\frac{Nb^2}{D^2}\right)^{\frac{1}{3\nu}}.
\end{equation}
Number of monomers per blob in the two cases are
\begin{equation}
g_I \sim \left(\frac{qE_ob}{k_BT}\right)^{\frac{-1}{1+\nu}}, \\
g_{II} \sim
\left(\frac{qE_ob}{k_BT}\frac{Nb^2}{D^2}\right)^{\frac{-1}{3\nu}}.
\label{g}
\end{equation}
Free energy per monomer behind the channel
is $\sim\frac{k_BT}{g_m}$.

Consider a polymer in each of the two cases that is going to
enter a channel of diameter $d$ on the wall. Entropic energy of
each monomer after entry into the channel is $\sim
\frac{k_BT}{g_{in}}$, in which,
$g_{in}=\left(\frac{d}{b}\right)^{1/\nu}$ is the number of
monomers per blob inside the channel. Electric energy change of
$n$ monomers after entering to the channel is $\sim -nqE_il$, in
which $E_i$ is the electric field strength inside the channel and
$l=\frac{n}{g_{in}}d=n(\frac{b}{d})^{1/\nu}d$ (see Fig.
\ref{fig1} (c)). Accordingly, total free energy change of the
polymer after entry of $n$ monomers is $\Delta F_m \sim k_BT
\left(\left(\frac{b}{d}\right)^{\frac{1}{\nu}}-\frac{1}{g_m}\right)n-n^2qE_i(\frac{b}{d})^{1/\nu}d$.
Free energy barrier height for the polymer entry, $\Delta U_m$,
is the maximum of $\Delta F_m$ with respect to $n$ and is given by
\begin{equation} \label{dF}
\frac{\Delta U_m}{k_BT}\sim
\left(\left(\frac{b}{d}\right)^{\frac{1}{\nu}}-\frac{1}{g_m}\right)^2
\frac{k_BT}{qE_id}\left(\frac{d}{b}\right)^{1/\nu}.
\end{equation}

In simulations for checking our theory, we model the polymer by
$N$ spherical monomers of diameter $\sigma$ connected by FENE
potential. Shifted and truncated Lennard-Jones potential is used
to model monomer-monomer and monomer-wall excluded volume
interactions and it is assumed that $E_o=E_i=E$. All our
simulations are performed with ESPResSo \cite{holm} using
Langevin thermostat to keep the temperature fixed.

To check our theoretical results on static behavior of the polymer
before its entry to the channel, we measure some variables in our
simulations by keeping the channel closed and letting the polymer
to fluctuate behind it. Lateral gyration radius of a polymer
confined between two planes of spacing $L$ is $R_{\parallel} \sim
N^{\frac{3}{4}}b \left(\frac{b}{L}\right)^{\frac{1}{4}}$
\cite{pressure}. For a fixed diameter of the confining cylinder,
$D=12\sigma$, exponents $\beta$ and $\gamma$ in relations
$L\propto N^{\beta}$ and $R_{\parallel}L^{\frac{1}{4}}\propto
N^{\gamma}$ are obtained via local slopes of the log-log plots of
$L$ and $R_{\parallel}L^{\frac{1}{4}}$ versus $N$. Results for two
different field strengths are shown in Fig. \ref{fig2} (a). For a
fixed value of $D$, short and long polymers correspond to cases
$I$ and $II$, respectively. With increasing $N$, crossover from
case $I$ with expected exponents $\beta_I=0$ and $\gamma_I=0.75$
to case $II$ with $\beta_{II}=\frac{1}{3\nu}\simeq 0.6$ and
$\gamma_{II}=\frac{1}{12\nu}\simeq 0.14$ can be seen in Fig.
\ref{fig2} (a). The value of $\alpha$ in relation $L\propto
E^{-\alpha}$ is also obtained from simulations for three
different polymer lengths. For both cases $I$ and $II$, it is
found that $\alpha\simeq 0.4$, in agreement with the theory.
Results on static behavior of the polymer is of importance in the
subject of polymer compression by external force in nanochannels
\cite{craighead2}. As an example and for comparison with
experiment, the change in the free energy of a polymer arising
from its confining in a narrow channel under an electric field
can be written from case $II$ of our theory as $\frac{\Delta
F}{k_BT}\sim\frac{N}{g_{II}}\propto N^{1.6}$. This is in excellent
agreement with experimental result of Ref. \cite{partitioning}.

\begin{figure}
\includegraphics[scale=0.55]{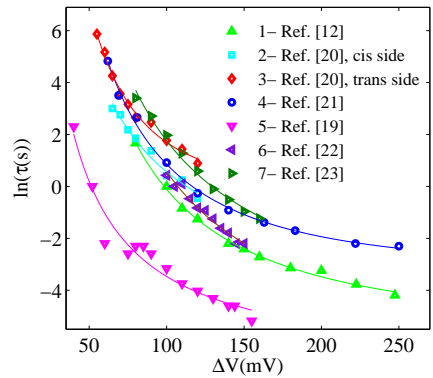}
\caption{ (color online) Logarithm of capture time versus $\Delta
V$ extracted from referenced articles (symbols) and fitted
equation of our theory, Eq. \ref{comp3} (solid lines).
All the data sets are from polymer translocation
experiments through $\alpha$-hemolysin channel. As it can be seen,
all data sets are followed well by the suggested function. Fit
parameters are shown in Table \ref{table1}.} \label{fig3}
\end{figure}

To check the main result of our theory, Eq. \ref{dF}, we measure
the barrier height for the polymer entry for each given set of $N$
and $E$ in both of cases $I$ and $II$. We fix the end monomer of
the polymer at the channel entry and leave the other monomers to
equilibrate under the applied electric field. Then we release the
polymer end to probe whether its entry is successful or no. We
obtain $-\frac{\Delta U}{k_BT}$ as the logarithm of successful
entry probability (fraction of at least $10^3$ simulations in
which the polymer entry is successful). Results for polymers of
different lengths at different values of the electric field
strength are shown in Fig. \ref{fig3} (b) and (c). As it can be
seen, there is a very good agreement between our theory and
simulation results in both of the two cases. In the insets of Fig.
\ref{fig2} (b) and (c), deviation of $\Delta U$ versus $E^{-1}$
from linear form in large intervals of $E$ in both cases, no
dependence of entry barrier on polymer length in case $I$
contrary to case $II$, and decrease of entry barrier with the
polymer length in case $II$ can be seen. These results are different
from predictions of the previous theories: linear decrease
of the barrier height with the applied voltage \cite{Chou-Grosberg} and its
dependence on the polymer length even for simple pores \cite{Muthukumar}. In previous studies,
dependence of the barrier height on $N$ is attributed to the restriction of the polymer end to the channel during
its entrance \cite{Kumar}. Because of the compression effect of the driving field which is taken into account in our work, only monomers in the first blob feel the restriction and its energy cost is independent of $N$.

One should note that the validity range of our theory in
both cases is $1\leq g_m \leq N$. $g_m>N$ means that the electric
field is too weak to has confining effect on the polymer behind
the channel and cannot overcome the entropic barrier. Consistent
with this description, in simulations with smaller polymers and
lower electric fields, we obtained $\alpha\simeq 0$ and
$\beta\simeq \nu$. $g_m<1$ means that the polymer loses most of
its entropy behind the channel and feels no entropic barrier on
its entry (as simulation results also show in Fig. 2 (b) and (c)).
Existence of two voltage limits, the one below which no polymer
entry happens and another one above which voltage dependence of the
capture time changes from exponential to linear function
\cite{capture2,expdata2} could be related to these limits.

In voltage driven polymer translocation experiments, one of the measurable quantities is the capture
time, $\tau$, at different voltage differences, $\Delta V$
 \cite{capture,capture2,barrier,expdata1,expdata2,expdata3,expdata4}.
The average capture time can be written from Van't Hoff-Arrhenius
law as $\tau = \tau_0 \exp\left(\frac{\Delta U}{k_BT} \right)$
\cite{barrier}, in which $\Delta U$ is the free energy barrier
height. $\tau_0^{-1}$ is the frequency of polymer fluctuations
behind the channel when entry of folded polymer is possible
\cite{park} or frequency of the polymer ends reaching to the
channel entrance when entry of folded polymer is not possible.
From Van't Hoff-Arrhenius law and eq. \ref{dF} of the theory, and
considering that $1/(1+\nu)\simeq 1/3\nu\simeq 0.6$, the equation
\begin{equation} \label{comp3}
\ln \tau =\zeta + \left( \eta -\lambda
\left(\frac{qE_ib}{k_BT}\right)^{0.6}\right)^2\frac{k_BT}{qE_ib}
\end{equation}
can be used for fitting to experimental data by $\zeta$, $\eta$,
and $\lambda$ as the fit parameters, regardless of the similarity
of the channel entrance geometry to case $I$ or $II$ of the
theory. In equation \ref{comp3}, we have assumed $E_i\simeq
\frac{\Delta V}{\ell}$ and $E_o=\delta E_i$, where $\ell$ is the
\begin{table}
\caption{\label{table1}Parameters $\zeta$, $\eta$, and $\lambda$
obtained from fitting of Eq. \ref{comp3} to seven data sets shown
in Fig. \ref{fig3}.}
\begin{ruledtabular}
\begin{tabular}{ccccc}
&Experimental data  &$\zeta$ & $\eta$
&$\lambda$ \\
\hline
1 &ssDNA, cis side (Ref. \cite{capture2})& -5.42 & 3.14 & 0.60 \\
2 &ssDNA, cis side (Ref. \cite{barrier})& -4.33 & 2.38 & 0 \\
3 &ssDNA, trans side (Ref. \cite{barrier})& 0.27 & 3.22 & 1.77 \\
4 &ssDNA, cis side (Ref. \cite{expdata1}) & -3.01 & 3.11 & 0.90\\
5 &DS, cis side (Ref. \cite{expdata2})& -7.09 & 2.35 & 0.01  \\
6 &DS, trans side (Ref. \cite{expdata3})& -6.13 & 3.85 & 0.51  \\
7 &PSS, cis side (Ref. \cite{expdata4})& -6.01 & 2.83 & 0  \\
\end{tabular}
\end{ruledtabular}
\end{table}
channel length and $\delta< 1$ (there is no satisfactory
knowledge about values and spatial functionality of $E_i$ and
$E_o$). Also, $b$ and $q$ are obtained by course-graining the
polymer such that  the size of effective spherical monomers is
equal to the polymer diameter. For example, for translocation of
single stranded DNA (ssDNA) through $\alpha$-hemolysin channel at
the room temperature, we use $\ell=10nm$, $b=1nm$, and $q=3e$
($e$ is the electron charge), which results in
$\frac{qE_ib}{k_BT}\simeq 0.01 \Delta V(mV)$. It should be noted
that from the Van't Hoff-Arrhenius law $\zeta = \ln\tau_0$ and
from Eqs. \ref{g} and \ref{dF}, $\eta\sim \left( \frac{d}{b}
\right)^{\frac{1}{2}-\frac{1}{2\nu}}$ for both of the two cases,
$\lambda\sim \left( \frac{d}{b}
\right)^{\frac{1}{2}+\frac{1}{2\nu}}\delta^{0.6}$ for case $I$
and $\lambda\sim \left( \frac{d}{b}
\right)^{\frac{1}{2}+\frac{1}{2\nu}}\delta^{0.6}
\left(\frac{Nb^2}{D^2} \right)^{0.6}$ for case $II$.

In Fig. \ref{fig3}, $\ln \tau$ versus $\Delta V$ reported for
translocation of three flexible polymers, ssDNA, dextran sulfate
(DS), and poly(styrenesulfonic acid) (PSS), from cis and trans
sides of $\alpha$-hemolysin channel and fitted function of Eq.
\ref{comp3} are shown. The values of fitting parameters are shown
in Table \ref{table1}. $\zeta$ in Eq. \ref{comp3} depends only on
$\tau_0$ which is out of interest here. $\eta$ depends  on the
ratio of the channel diameter to the monomer size and as all the
seven data sets shown in Fig. \ref{fig3} are for experiments with
the same channel, $\alpha$-hemolysin, similar values of $\eta$ is
reasonable. The most important parameter which is a measure of
the squeezing effect of the electric field on the polymer before
its entry to the channel is $\lambda$. For data sets 2, 5 and 7,
where polymer is added to the cis side and applied voltages are
relatively low (up to 160 $mV$), $\lambda$ is negligible. This is
because of the weak electric fields which results in small
squeezing effects. Nonzero value of $\lambda$ with similar
applied voltages for the trans side (data sets 3 and 6), could be
because of the smaller diameter of the channel in this side which
causes the electric field in the vicinity of the channel entrance
to be stronger and the parameter $\delta$ to be larger
\cite{electric-field}. It should be noted that according to ref.
\cite{capture2}, compression of the polymer inside the cis side
vestibule of $\alpha$-hemolysin channel happens very rarely. So
entry of the polymer from both the cis and trans sides is similar
to case $I$ of the theory (and not the case $II$). For the cis
side, data sets 1 and 4, applied voltages are higher (up to 280
$mV$) which makes the squeezing effect more considerable and the
value of $\lambda$ nonzero.

In summary, free energy barrier height for electric field driven
polymer entry into channels of two different entrance geometries
has been calculated theoretically and by MD simulations. The
barrier height dependence on the electric field strength and the
polymer length, the effect of lateral confinement of the polymer
before its entry to the channel on these dependencies and
relevance of results with experiment have been discussed.

We are grateful to Mohammad R. Kolahchi for useful comments.

\end{document}